\documentclass{ESI_Proc}

\contact[spohn@ma.tum.de]{Herbert Spohn\\
Zentrum Mathematik and Physik Department, TU M\"{u}nchen,\\
Boltzmannstr. 3, D-85747 Garching, Germany\\}

\numberwithin{equation}{section}

\theoremstyle{definition}

\usepackage{epsfig}
\usepackage{wrapfig}
\usepackage{psfrag}
\usepackage{subfigure}
\usepackage{nameref}
\usepackage[ansinew]{inputenc}
\usepackage{txfonts}

\title[Weakly  nonlinear wave equations]
{On the Boltzmann Equation for Weakly
 Nonlinear Wave Equations}

\author[Herbert Spohn]{Herbert Spohn\\
Zentrum Mathematik and Physik Department, TU M\"{u}nchen,\\
Boltzmannstr. 3, D-85747 Garching,  Germany\\}

\begin{document}

\maketitle

\begin{abstract}
We explain how the kinetic theory of L. Boltzmann is applied to
weakly nonlinear wave equations.
\end{abstract}

\section*{Introduction}

Kinetic Theory is an everlasting contribution of Ludwig Boltzmann to
physics. While he devised the theory to understand the dynamics of
dilute gases, the method as such is much more general and has been
applied with great success in many areas. A more recent example is
the flow of granular media. Their defining property are inelastic
collisions. Thus a natural, in fact highly informative, approach is
to follow the route of kinetic theory, adjusting for the appropriate
collision mechanism \cite{BP}.

In the early days of kinetic theory it was a convoluted process to
understand that the Boltzmann equation cannot possibly hold for
every mechanical initial condition. But to set forward the
proposition that the Boltzmann equation becomes exact in a
particular limiting procedure, now called the Boltzmann-Grad limit
\cite{Gr}, took almost 80 years counted from Boltzmann's 1872 paper.
In this limit the sometimes painfully vague notions of
``overwhelming probability'' and the like acquire a definite
meaning. For example, the set of exceptional initial phase points
has a measure (w.r.t. the initial measure on phase space) which
tends to zero in the Boltzmann-Grad limit. Grad's proposition was
proved as a mathematical theorem by Lanford \cite{La}, see
\cite{CIP,Sp1} for more expanded discussions.

Given the example of low density gases, one might wonder whether
similar ideas are applicable to further microscopic systems. The
Landau equation for weakly interacting particles and the Vlasov
equation for weak, long range forces are examples from plasma
physics. After the advent of quantum mechanics, an obvious task was
to also adjust kinetic theory. For electrons, satisfying Fermi-Dirac
statistics, the generalization was accomplished by Nordheim
\cite{No}, for the quantized wave equation (phonons) by Peierls
\cite{Pe}, and for dilute quantum gases by Uehling and Uhlenbeck
\cite{UU}.

The derivation of the Boltzmann equation for interacting quantum
particles is well covered in recent articles \cite{ErSaY,BeCa,Sp2}. Here
I summarize the current status for weakly nonlinear wave equations
with random initial data. On the surface this problem looks simpler
than weakly interacting quantum particles, since there is no need
for operators, Fock spaces, and the like. On the other hand, at
least on the level of time-dependent perturbation theory, the wave
equation and quantum systems have similar Feynman diagrams. In
particular cases the wave diagrams are a proper subset of the set of
all quantum diagrams. As a consequence the limiting kinetic
equations differ only minimally. For the kinetic description a
dividing line seems to be point particles versus wave equation, the
latter case being much less understood because of the nonlocal
interaction once the dynamics is expressed in terms of multi-point
Wigner functions.

Writing down a kinetic equation always involves some modeling
aspects. The available experience provides sufficient guidance as to
what would constitute an acceptable kinetic equation. There is then
no point in waiting for a firm microscopic derivation. Rather more
urgently is to work out predictions from the kinetic equation, which
by itself is a demanding endeavour. In fact, a kinetic equation to
be exact in a particular limit might very well be exceptional. A
prominent example are lattice Boltzmann algorithms, which are used
widely to simulate flows in complicated geometries and close to
reacting surfaces. For such algorithms the Boltzmann equation
remains as an approximation, although a very powerful one.

To give a brief outline, in the following section we explain, what I
call, the kinetic framework. Physicists tend to be more
computationally oriented and the underlying statistical properties
of the microscopic system are rarely spelled out in full detail. But
on a conceptual level the kinetic framework is most useful and,
assuming it to hold, the kinetic equation is obtained rather easily.
We emphasize that the kinetic framework is not restricted to a
particular model. It is a general structure, which can be used
whenever a splitting into a ``free'' part and an appropriately small
perturbation is meaningful. In Section \ref{sec.3} we write down the
phonon Boltzmann equation in case of a quartic potential. Some of
the basic properties of this equation are discussed in Section
\ref{sec.4}. Perhaps somewhat unexpected, the kinetic approach
yields interesting dynamical information even for an anharmonic
chain. In this light we revisit the famous Fermi-Pasta-Ulam problem.

\section{The kinetic framework}\label{sec.2}
\setcounter{equation}{0}

We plan to study the scalar wave equation in three space dimensions
with a cubic nonlinearity of the form
\begin{equation}\label{2.1}
\frac{\partial^2}{\partial
t^2}u=(\Delta-\omega^2_0)u-\sqrt{\lambda}u^3\,.
\end{equation}
Here $u:\mathbb{R}\times\mathbb{R}^3\to\mathbb{R}$ is the wave
field, $\Delta$ the Laplacian, and $\lambda>0$, but small. The
initial data are random and of finite energy. As will be explained,
the total energy is taken to be of order $\lambda^{-3}$ and the
support of $u$ of linear extent $\lambda^{-1}$. To avoid the issue
of ultraviolet divergencies, Eq. (\ref{2.1}) is discretized
spatially through the standard grid $\mathbb{Z}^3$. It is important
to also allow for a general dispersion relation of the linear part.
Physically, the discretized wave equation describes the lattice
dynamics of dielectric crystals, see \cite{Lei,Zi,Gu} for systematic
expositions. In this context the scalar equation would correspond to
a one-band approximation.

Fourier transform will be a convenient tool. Let
$\mathbb{T}^3=[-\tfrac{1}{2},\tfrac{1}{2}]^3$ be the first Brillouin
zone of the lattice dual to $\mathbb{Z}^3$. For
$f:\mathbb{Z}^d\to\mathbb{R}$ its Fourier transform, $\widehat{f}$,
is defined by
\begin{equation}\label{2.2}
\widehat{f}(k)=\sum_{x\in\mathbb{Z}^3}e^{-i 2\pi k\cdot x} f_x\,.
\end{equation}
Here $k\in\mathbb{T}^3$ and $\widehat{f}$ extends periodically to a
function on $\mathbb{R}^3$. The inverse Fourier transform is given
by
\begin{equation}\label{2.3}
f_x=\int_{\mathbb{T}^3} dk e^{i 2\pi k\cdot x} \widehat{f}(k)\,.
\end{equation}

For $x\in\mathbb{Z}^3$ the displacement away from $x$ is denoted by
$q_x\in\mathbb{R}$ and the corresponding momentum by
$p_x\in\mathbb{R}$. In the harmonic approximation the interaction
potential is
\begin{equation}\label{2.4}
U_\mathrm{harm}(q)=\frac{1}{2}\sum_{x,y\in\mathbb{Z}^3}\alpha(x-y)q_x
q_y\,.
\end{equation}
The elastic constants $\alpha(x)$ satisfy
\begin{equation}\label{2.5}
\alpha(x)=\alpha(-x)\quad\mathrm{and} \quad |\alpha(x)|\leq \gamma_0
e^{-\gamma_1|x|}
\end{equation}
for suitable $\gamma_0,\gamma_1>0$. Mechanical stability requires
\begin{equation}\label{2.6}
\widehat{\alpha}(k)\geq 0\,.
\end{equation}
In fact, we assume $\widehat{\alpha}(k)>0$ for $k\neq 0$. The case
$\widehat{\alpha}(k)\simeq k^2$ for small $k$ is refered to as
`acoustical' while $\widehat{\alpha}(0)>0$ is `optical'. The
dispersion relation of the harmonic part is
\begin{equation}\label{2.7}
    \omega(k)=\sqrt{\widehat{\alpha}(k)}\,.
\end{equation}

Following (\ref{2.1}) as a concrete example, we consider a quartic
on-site potential. Then the full hamiltonian of the lattice dynamics
is given by
\begin{equation}\label{2.8}
H=\frac{1}{2}\sum_{x\in\mathbb{Z}^3}p^2_x+
\frac{1}{2}\sum_{x,y\in\mathbb{Z}^3}\alpha(x-y)q_x q_y+
\frac{1}{4}\sqrt{\lambda}\sum_{x\in\mathbb{Z}^3}q^4_x\,.
\end{equation}
The harmonic part of $H$ will be denoted by $H_\mathrm{ha}$,
$H=H_\mathrm{ha}+\sqrt{\lambda}V$. In a sense to be explained below,
$\sqrt{\lambda}V$ is a ``small'' perturbation of $H_\mathrm{ha}$.

We concatenate $q_x$ and $p_x$ into a single complex-valued field
$a(k)$ as
\begin{equation}\label{2.9}
a(k)=\frac{1}{\sqrt{2}}\Big(\sqrt{\omega(k)}\widehat{q}(k) + i
\frac{1}{\sqrt{\omega(k)}}\widehat{p}(k)\Big)
\end{equation}
with the inverse
\begin{equation}\label{2.10}
\widehat{q}(k)=\frac{1}{\sqrt{2}}\frac{1}{\sqrt{\omega(k)}}\big(a(k)+
a(-k)^\ast\big)\,,\;
\widehat{p}(k)=\frac{i}{\sqrt{2}}{\sqrt{\omega(k)}}\big(-a(k)+
a(-k)^\ast\big)\,.
\end{equation}
Then the hamiltonian reads
\begin{equation}\label{2.11}
H=H_\mathrm{ha}+\sqrt{\lambda}V\,,\quad
H_\mathrm{ha}=\int_{\mathbb{T}^3}dk \omega(k)a(k)^\ast a(k)\,,
\end{equation}
\begin{eqnarray}\label{2.12}
&&\hspace{-30pt}V=\frac{1}{4}\int_{\mathbb{T}^{12}}dk_1 dk_2 dk_3
dk_4 \delta(k_1+k_2+k_3+k_4)
\nonumber\\
&&\hspace{20pt}\times\prod^4_{j=1}\big(2\omega(k_j)\big)^{-1/2}
\big(a(k_j)+a(-k_j)^\ast\big)\,.
\end{eqnarray}
Hence the equations of motion for the $a$-field are
\begin{eqnarray}\label{2.13}
&&\hspace{-30pt}\frac{\partial}{\partial t}a(k,t)=-i\omega(k)a(k,t)
- i\sqrt{\lambda}\int_{\mathbb{T}^9}dk_1 dk_2 dk_3
\delta(k-k_1-k_2-k_3)\nonumber\\
&&\hspace{23pt}\times\big(2\omega(k)\big)^{-1/2}\prod^3_{j=1}\big(2\omega(k_j)\big)^{-1/2}
\big(a(k_j,t)+a(-k_j,t)^\ast\big)\,.
\end{eqnarray}
In particular for $\lambda=0$,
\begin{equation}\label{2.13a}
    a(k,t)=e^{-i\omega(k)t}a(k)\,.
\end{equation}

Let us now briefly recall the case of dilute gases. The central
quantity is the Boltzmann $f$-function, which is the number density
on one-particle phase space $\mathbb{R}^3\times\mathbb{R}^3$. It
changes in time by the free motion of particles and through pair
collisions,
\begin{equation}\label{2.14}
\frac{\partial}{\partial t}f_t+v\cdot\nabla_r
f_t=\mathcal{C}(f_t,f_t)\,.
\end{equation}
We do not write out the collision operator explicitly but note that
it is bilinear in $f_t$ and strictly local in the space-variable
$r$, since on the kinetic scale particles collide at the same point.
At low density the full particle statistics is close to Poisson,
jointly in positions and velocities, with an intensity given again
by the $f$-function. This double meaning of the $f$-function is a
source of conceptual confusion. In the Sto{\ss}zahlansatz one
assumes independent incoming velocities, thus the second meaning of
$f$, while the predictions of kinetic theory are based on a law of
large numbers, thus the first meaning of $f$. The Poisson statistics
appears naturally through the good statistical mixing properties of
the collisionless dynamics. In fact, at infinite volume, these are
the only translation invariant measures, which are invariant under
the collisionless dynamics and which have a finite number, energy,
and entropy per unit volume \cite{EySp}.

For wave equations the Poisson measure will be substituted by a
Gaussian measure, constrained to be locally invariant under the
dynamics generated by $H_\mathrm{ha}$. In the \textit{spatially
homogeneous} case, imposing time-stationarity and using the explicit
solution (\ref{2.13a}) for $\lambda=0$, the $\{a(k)$,
$k\in\mathbb{T}^3\}$ are then jointly Gaussian with mean zero and
covariance
\begin{equation}\label{2.15}
\langle a(k)^\ast a(k')\rangle=W(k)\delta(k-k')\,,
\end{equation}
\begin{equation}\label{2.16}
\langle a(k)^\ast a(k')^\ast\rangle=0\,,\quad \langle
a(k)a(k')\rangle=0\,,
\end{equation}
defining the power spectrum $W(k)$. Clearly, $W(k)\geq 0$.

For the \textit{spatially inhomogeneous} case the construction is
slightly more elaborate. Since the anharmonic potential is of order
$\sqrt{\lambda}$, the mean free path for phonons is of order
$\lambda^{-1}$. This is the scale on which spatial inhomogeneities
have to be imposed. Local stationarity still implies mean zero and
(\ref{2.16}). For the covariance (\ref{2.15}) we first prescribe the
limiting local power spectrum
$W:\mathbb{R}^3\times\mathbb{T}^3\to\mathbb{R}$ with $W\geq 0$ and a
rapid decay in position space. In analogy to the semiclassical limit
of the Schr\"{o}dinger equation we refer to $W$ as Wigner function.
A  \textit{locally stationary} Gaussian measure, denoted by
$\langle\cdot\rangle^{G,\lambda}$, is in fact a scale of probability
measures depending on $\lambda$. For each $\lambda$ the measure
$\langle\cdot\rangle^{G,\lambda}$ is Gaussian with
\begin{equation}\label{2.17}
\langle a(k)\rangle^{G,\lambda}=0\,,\quad\langle
a(k)a(k')\rangle^{G,\lambda}=0\,.
\end{equation}

Following Wigner the local power spectrum of the $a$-field is
defined through
\begin{equation}\label{2.18}
P^1(x,k)=2^{-3}\int_{(2\mathbb{T})^3}d\eta e^{i2\pi x\cdot\eta}
a(k-\eta/2)^\ast a(k+\eta/2)\,.
\end{equation}
We rescale the lattice to have lattice spacing $\lambda$ through the
substitution $x=\lambda^{-1}y$, $y\in(\lambda\mathbb{Z}/2)^3$, and
obtain the rescaled local power spectrum
\begin{equation}\label{2.19}
P^\lambda(y,k)=(\lambda/2)^3\int_{(2\mathbb{T}/\lambda)^3}d\eta
e^{i2\pi y\cdot\eta} a(k-\lambda\eta/2)^\ast a(k+\lambda\eta/2)\,.
\end{equation}
$P^\lambda$ is a random field. By definition, its average is the
one-particle Wigner function,
\begin{equation}\label{2.20a}
W^\lambda_1(y,k)=\langle P^\lambda(y,k)\rangle^{G,\lambda}
\end{equation}
and its variance the two-particle Wigner function,
\begin{equation}\label{2.20b}
W^\lambda_2(y_1,k_1,y_2,k_2)=\langle
P^\lambda(y_1,k_1)P^\lambda(y_2,k_2)\rangle^{G,\lambda}\,.
\end{equation}
With $\lfloor\cdot\rfloor_\lambda$ denoting modulo $\lambda$, the
scale of Gaussian measures $\langle\cdot\rangle^{G,\lambda}$ is
assumed to satisfy the pointwise limit
\begin{equation}\label{2.20}
\lim_{\lambda\to 0}W^\lambda(\lfloor r\rfloor_\lambda,k)=W(r,k)\,.
\end{equation}
Furthermore one has to require a law of large numbers for the power
spectrum $P^\lambda$ which can be expressed in the form
\begin{equation}\label{2.21a}
\lim_{\lambda\to 0}W^\lambda_2(\lfloor
r_1\rfloor_\lambda,k_1,\lfloor
r_2\rfloor_\lambda,k_2)=W(r_1,k_1)W(r_2,k_2)
\end{equation}
for $r_1\neq r_2$.

In general, the assumption of strict Gaussianity is too strong.
Because of the anharmonicity, unavoidably there will be small
errors. To allow for them we call an arbitrary sequence of
probability measures, $\langle\cdot\rangle^\lambda$, \textit{locally
stationary}, if the Gaussian property is gained only in the limit
$\lambda\to 0$. More precisely, for our model we may assume that all
odd moments of $\langle\cdot\rangle^\lambda$ vanish, since this
property is propagated in time. For the second moments $\langle
a(k)a(k')\rangle^\lambda$, $\langle a(k)^\ast a(k')\rangle^\lambda$
we form the rescaled one-particle Wigner function as in
(\ref{2.19}), (\ref{2.20a}). The rescaled Wigner function for
$\langle a(k)a(k')\rangle^\lambda$ is assumed to vanish as
$\lambda\to 0$, while the rescaled Wigner function for $\langle
a(k)^\ast a(k')\rangle^\lambda$ satisfies (\ref{2.20}).
(\ref{2.21a}) is imposed correspondingly. In addition we require
that the sequence of local measures, close to $\lambda^{-1}r$
[lattice units], converges to a Gaussian measure with covariance
(\ref{2.15}), (\ref{2.16}) and $W(k)$ replaced by $W(r,k)$.

 From
(\ref{2.20}) we infer that for a locally stationary measure it holds
\begin{equation}\label{2.21}
\langle H_\mathrm{ha}\rangle^{\lambda}=\int_{\mathbb{T}^3}dk
\omega(k)\langle a(k)^\ast
a(k)\rangle^{\lambda}=\mathcal{O}(\lambda^{-3})\,.
\end{equation}
The harmonic energy is extensive. On the other hand the average
anharmonic potential satisfies $\sqrt{\lambda}\langle
V\rangle^{\lambda}=\sqrt{\lambda}\mathcal{O}(\lambda^{-3})$, which
is small compared to $\langle H_\mathrm{ha}\rangle^{\lambda}$.

Our real goal is to understand the time evolution in the limit of
small $\lambda$. We impose the initial measure to be Gaussian and
locally stationary. As a simple, but useful first step we set $V=0$
and, for a short intermediate stretch only, consider the linear
dynamics generated by $H_\mathrm{ha}$. By linearity this dynamics
preserves the Gaussian property. Since the initial state has a slow
variation on the scale $\lambda^{-1}$, the limit $\lambda\to 0$ for
the one-particle Wigner function is a particular case of the
semiclassical limit for a linear wave equation. Thus the appropriate
time scale is also of order $\lambda^{-1}$ and on that scale
\begin{equation}\label{2.22}
\lim_{\lambda\to 0} W^\lambda(\lfloor
r\rfloor_\lambda,k,\lambda^{-1}t)=W(r,k,t)\,.
\end{equation}
\big(On the left hand side $\lambda^{-1}t$ is the time in
microscopic units, the units of (\ref{2.13}), while on the right
hand side $t$ refers to kinetic times, the units for
(\ref{2.14})\big). The limit Wigner function is governed by the
transport equation
\begin{equation}\label{2.23}
\frac{\partial}{\partial
t}W(r,k,t)+\frac{1}{2\pi}\nabla_k\omega(k)\cdot\nabla_r
W(r,k,t)=0\,,
\end{equation}
which corresponds to the free motion of fictiuous particles, the
phonons, with kinetic energy $\omega$. Of course, (\ref{2.23}) has
to be solved with the initial condition $W(r,k,0)=W(r,k)$.

We refer to \cite{Mie,LukTeu} for a complete coverage of the
semiclassical limit for lattice dynamics with an arbitrary unit
cell. Random initial data are studied in \cite{DuSp}, where only
rather mild mixing conditions on the initial measure
$\langle\cdot\rangle^\lambda$ are imposed. Let
$\langle\cdot\rangle^\lambda_t$ be the measure at time $t$ as
evolved according to the dynamics generated by $H_\mathrm{ha}$. It
is proved that for any $t>0$ the sequence of measures
$\langle\cdot\rangle^\lambda_{\lambda^{-1}t}$ is locally stationary
with a Wigner function satisfying the transport equation
(\ref{2.23}). In this sense local stationarity is dynamically
generated. Such results require, as does the kinetic limit, that
phonons propagate with nonzero velocity. To say one has to demand
that $\nabla_k\omega\neq 0$ a.s.. By our assumption, $\omega^2$ is
real analytic on $\mathbb{T}^3$ and one only has to exclude the case
$\omega=$ \textit{const}.. Nevertheless, since momentum space is
$\mathbb{T}^3$, there will always be submanifolds where
$\nabla_k\omega=0$, which sets an extra technical difficulty.

We return to the case of interest, namely adding the anharmonicity
$\sqrt{\lambda}V$. Then, if as before the initial measure is
required to be Gaussian and locally stationary, the time-evolved
measure $\langle\cdot\rangle^{G,\lambda}_t$ is no longer Gaussian.
On the other hand, the linear dynamics does not tolerate deviations
from Gaussianity, as demonstrated in \cite{DuSp}, and the strength
of the anharmonicity is taken to be small. Thus the family
$\langle\cdot\rangle^{G,\lambda}_{\lambda^{-1}t}$ is still locally
stationary. Of course now $W(k)$ depends on $r$ and $t$ [kinetic
scale]. The precise strength of the anharmonic on-site potential is
adjusted such that its effect is of order 1 for the scale on which
(\ref{2.23}) holds. More physically speaking, the interaction
strength is such that the phonon mean free path is of order
$\lambda^{-1}$ [lattice units]. Thus one expects the limit
(\ref{2.22}) still to be valid. Only the transport equation picks up
a term resulting from the collisions between phonons.

\begin{figure}[h]
\begin{center}
\begin{psfrags}
\psfrag{SLa}[][][1]{$\mathsf S^\lambda$}
\psfrag{t0}[][][1]{$t=0$}
\includegraphics[height=4cm]{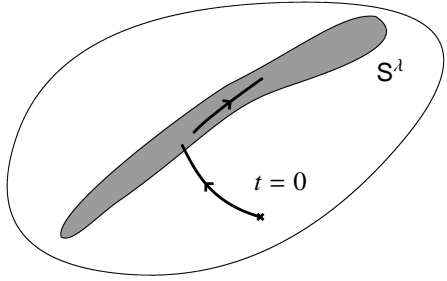}
\caption{Time evolution of a probability measure under kinetic scaling.}
\end{psfrags}
\end{center}
\end{figure}

We summarize the kinetic framework by means of a somewhat schematic
diagram. The encircled set is supposed to represent the set of
``all'' probability measures for the random field $a(k)$. We fix the
interaction strength as $0<\lambda\ll 1$. The manifold-like set
denoted by $\textsf{S}^\lambda$ is the set of all locally stationary
measures up to some $\lambda$-dependent scale of precision. The
hamiltonian dynamics generates a flow on the space of measures. The
set $\textsf{S}^\lambda$ is attractive, in the sense that fairly
rapidly, times of order 1, the measure will be close to
$\textsf{S}^\lambda$. On $\textsf{S}^\lambda$ the time evolution is
slow with changes on the kinetic scale $\mathcal{O}(\lambda^{-1})$
and the defining Wigner function is governed by a kinetic equation.
For $t\gg\lambda^{-1}$ further dynamical phenomena set in which
cannot be captured through the kinetic framework.

\section{The Boltzmann equation}\label{sec.3}
\setcounter{equation}{0}

The kinetic framework provides us with a tool for an educated guess
on the form of the collision operator. An important check will be
that the resulting equation has the physically correct stationary
solutions and satisfies an $H$-theorem for the total entropy,
respectively a local semi-conservation law with a positive entropy
production.

A more demanding issue is to support the educated guess by
mathematical arguments which establish the assumed existence of
limits and local stationarity.

Let $\langle\cdot\rangle^{G,\lambda}_t=\langle\cdot\rangle_t$ denote
the measure at time $t$ under the dynamics generated by $H$ with
$\langle\cdot\rangle_0=\langle\cdot\rangle^{G,\lambda}$. The
two-point function satisfies
\begin{equation}\label{3.1}
\frac{d}{dt}\langle a(p)^\ast a(q)\rangle_t =
i\big(\omega(p)-\omega(q)\big)\langle a(p)^\ast a(q)\rangle_t
+\sqrt{\lambda} F(q,p,t)\,.
\end{equation}
$F$ is cubic in the $a$-field. We integrate (\ref{2.13}) in time,
symbolically
\begin{equation}\label{3.2}
a(t)= e^{-i\omega t} a+\sqrt{\lambda}\int^t_0 e^{-i\omega(t-s)}
a(s)^3\,,
\end{equation}
and insert (\ref{3.2}) in the expression for $F$. There are then
cubic terms, proportional to $\sqrt{\lambda}$, which vanish upon
averaging, and sixtic terms, proportional to $\lambda$. At this
point we use, as argued, that locally the measure is approximately
Gaussian and factorize the 6-point function according to the
Gaussian rule for moments. We also rescale space-time by
$\lambda^{-1}$ and switch to Wigner function coordinates. In this
form one can take the limit $\lambda\to 0$. The first term on the
right of (\ref{3.1}) yields the free flow while $\sqrt{\lambda}F$
results in the collision operator with a cubic nonlinearity.

Details of such computations can be found in \cite{ErSaY,Sp3}. Here
we only quote the result, which is the Boltzmann equation
\begin{equation}\label{3.3}
\frac{\partial}{\partial t}
W(r,k,t)+\frac{1}{2\pi}\nabla_k\omega(k)\cdot\nabla_r W(r,k,t)
=\mathcal{C}(W(r,t))(k)\,.
\end{equation}
The collision operator is local in $r,t$, which our notation is
supposed to indicate. $\mathcal{C}$ is a nonlinear functional of
$k\mapsto W(r,k,t)$ at fixed $r,t$. For the quartic on-site
potential, $\mathcal{C}$ is cubic and defined through
\begin{eqnarray}\label{3.4}
&&\hspace{-48pt}\mathcal{C}(W)(k)= 12\pi
\sum_{\sigma_1,\sigma_2,\sigma_3=\pm 1}\int_{\mathbb{T}^9} d k_1 d
k_2 d k_3 (16 \omega\omega_1\omega_2\omega_3)^{-1}\nonumber\\
&&\hspace{36pt}\times\delta
(k+\sigma_1k_1+\sigma_2k_2+\sigma_3k_3)\delta
(\omega+\sigma_1\omega_1+\sigma_2\omega_2+\sigma_3\omega_3)\nonumber\\
&&\hspace{36pt} \times\big(W_1 W_2
W_3+W(\sigma_1W_2W_3+W_1\sigma_2W_3+W_1W_2\sigma_3)\big)\,,
\end{eqnarray}
where we use the shorthand $W_j=W(k_j)$, $\omega_j=\omega(k_j)$,
$j=1,2,3$. The term proportional to $W$ is the loss term. It has no
definite sign. On the other hand the gain term $W_1W_2W_3$ is always
positive, ensuring that the positivity of the initial Wigner
function is propagated in time.

As a historical parenthesis we remark that in his seminal paper
Peierls uses a very different reasoning. The current method goes
back to the mid-fifties where Green's function techniques were
developed systematically along the lines set forth by Quantum Field
Theory. The textbook expositions of the phonon Boltzmann equation
are not so satisfactory, at least in solid state physics. For the
quantized theory Fermi's golden rule is applied in a spatially
homogeneous, discrete mode context. While this gives the correct
collision operator, the `why' remains obscure. Classical lattice
dynamics is conceived only as a $\hbar\to 0$ limit of the quantum
theory. The situation is more favorable in wave turbulence, where
the starting point is a weakly nonlinear system of wave equations in
hamiltonian form \cite{Za}. Statistical properties are accessible
and the validity of Gaussian statistics is an important issue
\cite{Jan,CLN}.

At present, we are very far away from a complete mathematical proof
of the existence of the limit in Eq. (\ref{2.22}) together with the
property that the limiting Wigner function is the solution to
(\ref{3.3}),  (\ref{3.4}) with initial conditions $W(r,k,0)=W(r,k)$.
Apparently, the only available technique is to expand with respect
to $\sqrt{\lambda}V$ into a time-dependent perturbation series. Each
term of the series can be represented symbolically as a Feynman
diagram, which represents an oscillatory integral. In the limit
$\lambda\to 0$ most diagrams vanish (subleading diagrams) and a few
do not vanish (leading diagrams). The sum over all leading diagrams
yields the time-dependent perturbation series of the Boltzmann
equation, where the free flow is integrated and the collision term
is regarded as perturbation. The conjectured division into leading
and subleading diagrams is explained in \cite{HS}.

The program outlined has been carried out only fragmentarily. The
first order, proportional to $\sqrt{\lambda}$, vanishes. To second
order one expects the limit
\begin{equation}\label{3.5}
\int^t_0 ds e^{L(t-s)}\mathcal{C}(e^{Ls}W)\,,
\end{equation}
where $(e^{Lt}W)(r,k)=W(r-(2\pi)^{-1}\nabla_k\omega(k)t,k)$. While
such limit could be established at the level of generality discussed
here, even this point has not be accomplished. Closest to the goal
comes the very careful analysis of Ho and Landau \cite{HoL}, who
study the same problem for a weakly interacting Fermi liquid on
$\mathbb{Z}^3$ with nearest neighbor hopping and obtain the analogue
of (\ref{3.5}). Benedetto \textit{et al}. \cite{Be1,Be2} consider a
weakly interacting quantum fluid in $\mathbb{R}^3$ with the usual
kinetic energy $p^2/2m$. In our context this would correspond to the
nonlinear Schr\"{o}dinger equation, $m=1$,
\begin{equation}\label{3.6}
i \frac{\partial}{\partial t}\psi(x,t)=\big(-\tfrac{1}{2}\Delta_x
+\sqrt{\lambda}V_\psi(x,t)\big)\psi(x,t)
\end{equation}
with the effective potential
\begin{equation}\label{3.7}
V_\psi(x,t)=\int_{\mathbb{R}^3}dy \vartheta(x-y) |\psi(y,t)|^2\,.
\end{equation}
$\vartheta$ smoothens the interaction on a microscopic scale. The
role of the $a$-field is taken over by the Fourier transform
$\widehat{\psi}$. The Wigner function $W^\lambda(t)$ is precisely
the standard Wigner function for the Schr\"{o}dinger equation at
time $\lambda^{-1}t$ averaged over the initial Gaussian distribution
of the $\psi$-field. The Boltzmann equation for the nonlinear
Schr\"{o}dinger equation (\ref{3.6}) has the same overall structure
as (\ref{3.3}), (\ref{3.4}). More precisely, now $k\in\mathbb{R}^3$
and
\begin{equation}\label{3.7a}
\frac{\partial}{\partial t}W(r,k,t)+k\cdot \nabla_r
W(r,k,t)=\mathcal{C}_{\mathrm{NS}}(W(r,t))(k)
\end{equation}
with the collision operator
\begin{eqnarray}\label{3.8}
&&\hspace{-42pt}\mathcal{C}_{\mathrm{NS}}(W)(k_1)= 12\pi
\int_{\mathbb{R}^9} d k_2 d k_3 d k_4
|\widehat{\vartheta}(k_1-k_2)|^2
2\delta(k^2_1+k^2_2-k^2_3-k^2_4)\nonumber\\
&&\hspace{-5pt}\times \delta(k_1+k_2-k_3-k_4)
\big(W_2W_3W_4-W_1(W_2W_3+W_2W_4-W_3W_4)\big)\,.
\end{eqnarray}

The Feynman diagrams of the nonlinear Schr\"{o}dinger equation are a
proper subset of those investigated in \cite{Be1}. Using their
result, one concludes that in the limit $\lambda\to 0$ the second
order expansion term is given by (\ref{3.5}) with the appropriate
adjustments for $L$ and $\mathcal{C}$.

\section{Stationary solutions, entropy production}\label{sec.4}
\setcounter{equation}{0}

For dilute gases the $f$-function derived from the thermal
equilibrium distribution of the $N$-particle system is a stationary
solution of the kinetic equation. If one identifies the entropy with
the logarithm of phase space volume associated to some given
$f$-function on one-particle phase space, then the entropy
functional turns out to be
\begin{equation}\label{4.1a}
S(f)=-\int dv f(v)\log f(v)\,,
\end{equation}
in units where Boltzmann's constant $k_\mathrm{B}=1$. $S(f)$ is
increasing in time and constant if and only if $f$ is Maxwellian (at
least in the spatially homogeneous case). Thus the reader may wonder
whether the phonon Boltzmann equation has similar properties.

To determine the collision rule for phonons, one has to solve the
conservation laws of energy and momentum, see Eq (\ref{3.4}). For
pair  collisions, i.e. $\sum^3_{j=1}\sigma_j=-1$, this amounts to
\begin{equation}\label{4.1}
\omega(k_1)+\omega(k_2)=\omega(k_3)+\omega(k_1+k_2-k_3)
\end{equation}
and for three phonons mergers to
\begin{equation}\label{4.2}
\omega(k_1)+\omega(k_2)+\omega(k_3)=\omega(k_1+k_2+k_3)\,,
\end{equation}
where both equations implicitly define $k_3=k_3(k_1,k_2)$. $k_4$ is
obtained from momentum conservation as $k_4=k_1+k_2-k_3$ mod 1,
resp. $k_4=k_1+k_2+k_3$ mod 1. Mostly one has to work with this
implicit definition and only in very exceptional cases an explicit
collision formula is available. An instructive example is the case
one space dimension. Mechanical particles would merely exchange
labels and the Boltzmann collision operator vanishes. On the other
hand, for phonons in one dimension with an $\omega$ derived from
nearest neighbor couplings, (\ref{4.1}) has a nondegenerate
solution, thus providing real phonon collisions \cite{Per,ALS}. In
principle, it may happen that the energy-momentum conservation laws
have no solution at all. Then the collision term vanishes. In case
of several solutions one has to sum over all of them. Note that in
(\ref{3.4}) energy conservation cannot be satisfied for the term
with $\sigma_j=1$, $j=1,2,3$. This term had been added only to have
a more symmetric looking expression. A further instructive example
are nearest neighbor couplings, for which the dispersion relation
reads
\begin{equation}\label{4.3}
\omega(k)=\Big(\omega^2_0 + 2 \sum^3_{j=1}\big(1-\cos(2\pi
k^j)\big)\Big)^{1/2}\,,\quad k=(k^1,k^2,k^3)\in \mathbb{T}^3\,.
\end{equation}
Then (\ref{4.2}) has no solution and collision processes where three
phonons merge into one, and their time reversal, are forbidden. As
for dilute gases, there are only number conserving pair collisions.
Such a property is stable under small perturbations of $\omega$. It
also holds for the nonlinear wave equation (\ref{2.1}) for which
$\omega(k)=|k|$, $k\in\mathbb{R}^3$.

The equilibrium measure for the harmonic part is the Gaussian
$Z^{-1}\exp [-\beta H_\mathrm{har}]$ with $Z$ the normalizing
partition function and $\beta$ the inverse temperature, $\beta>0$.
The anharmonic potential can be neglected in the kinetic limit. The
corresponding Wigner function, $W_\beta$, is given by
\begin{equation}\label{4.4}
W_\beta(k)=\frac{1}{\beta\omega(k)}\,,
\end{equation}
in the limit of an infinitely extended lattice, compare with
(\ref{2.15}). Using (\ref{3.4}), it follows that
\begin{eqnarray}\label{4.5}
&&\hspace{-18pt}\mathcal{C}(W_\beta)(k)= \tfrac{3\pi}{4}
\sum_{\sigma_1,\sigma_2,\sigma_3=\pm 1}\int_{\mathbb{T}^3} d k_1 d
k_2 d k_3 (\omega\omega_1\omega_2\omega_3)^{-1}\delta
(\omega+\sigma_1\omega_1+\sigma_2\omega_2+\sigma_3\omega_3)\nonumber\\
&&\hspace{25pt}\times\delta
(k+\sigma_1k_1+\sigma_2k_2+\sigma_3k_3)\beta^{-3}
(\omega\omega_1\omega_2\omega_3)^{-1}
(\omega+\sigma_1\omega_1+\sigma_2\omega_2+\sigma_3\omega_3)\nonumber\\
&&\hspace{26pt}=0\,,
\end{eqnarray}
as expected. The issue whether there are further stationary
solutions will be discussed below.

Following Boltzmann, to define the entropy one first has to identify
a family of macroscopic observables. If for simplicity, and in fact
for the remainder of this section, we restrict ourselves to the
spatially homogeneous situation, then in spirit the macrovariables
are the phonon numbers $a(k)^\ast a(k)$, $k\in\mathbb{T}^3$. To be
more precise one has to employ the limit procedure of Boltzmann, the
only difference being that here we have to take into account the
harmonic couplings. We partition the torus $\mathbb{T}^3$ into cubes
$\Delta_j$ of side length $\delta$, $j=1,\ldots,M^3$, $\delta M=1$.
We consider a finite lattice volume, $\ell^3$, which is equivalent
to discretizing the torus $\mathbb{T}^3$ to $(\mathbb{T}_{\ell})^3$
with grid spacing $\ell^{-1}$. The proper macrovariables are then
\begin{equation}\label{4.6}
H_j=\sum_{k\in\Delta_j\cap(\mathbb{T}_{\ell})^3} a(k)^\ast
a(k)\,,\quad j=1,\ldots, M^3\,.
\end{equation}
For a given Wigner function $W$ the cell in phase space at precision
$\ell^3\varepsilon$ is defined by the conditions
\begin{equation}\label{4.7}
\{(q,p)\in{(\mathbb{R}^2)}^{{\ell}^3}|\,\ell^3(e_j-\varepsilon)\leq
H_j(q,p)\leq \ell^3(e_j+\varepsilon)\,,\quad j=1,\ldots,M^3\}
\end{equation}
with
\begin{equation}\label{4.8}
e_j=\delta^3\int_{\Delta_j}dk W(k)\,.
\end{equation}
The entropy $S(W)$ of $W$ is defined as $\ell^{-3}$ times the
logarithm of the Lebesgue measure of the set in (\ref{4.7}) upon
taking limits in the following order: $\ell\to\infty$,
$\varepsilon\to 0$, $\delta\to 0$. The net result is
\begin{equation}\label{4.9}
S(W)=\int_{\mathbb{T}^3} dk \log W(k)\,,
\end{equation}
up to a constant independent of $W$.

Before computing the rate of change of the entropy let us introduce
the notion of a collisional invariant. This is a function
$\psi:\mathbb{T}^3\to\mathbb{R}$ which satisfies either one of the
following functional equations.\smallskip\\
(i) (\textit{pair collisions})
\begin{equation}\label{4.10}
\psi(k_1)+\psi(k_2)=\psi(k_3)+\psi(k_1+k_2-k_3)
\end{equation}
on the set $\{(k_1,k_2,k_3)\in\mathbb{T}^9\,|\,\omega(k_1)+
\omega(k_2)=\omega(k_3)+\omega(k_1+k_2-k_3)\}$.\smallskip\\
(ii) (\textit{three phonons merger})
\begin{equation}\label{4.11}
\psi(k_1)+\psi(k_2)+\psi(k_3)=\psi(k_1+k_2+k_3)
\end{equation}
on the set $\{(k_1,k_2,k_3)\in\mathbb{T}^3\,|\,\omega(k_1)+
\omega(k_2)+\omega(k_3)=\omega(k_1+k_2+k_3)\}$.

If $W$ evolves according the spatially homogeneous kinetic equation
(\ref{3.3}), then
\begin{eqnarray}\label{4.12}
&&\hspace{-22pt}\frac{d}{dt}S(W)=\int_{\mathbb{T}} dk_1
W(k_1)^{-1}\mathcal{C}(W)(k_1)\nonumber\\
&&\hspace{-8pt}=12\pi\sum_{\sigma_1,\sigma_2,\sigma_3,\sigma_4=\pm
1}\int_{\mathbb{T}^4} d k_1 d k_2 d k_3 d k_4(16
\omega_1\omega_2\omega_3\omega_4)^{-1}\delta
\big(\sum^4_{j=1}\sigma_j \omega_j\big)\delta
\big(\sum^4_{j=1}\sigma_j k_j\big)\nonumber\\
&&\hspace{14pt}\times W^{-1}_1 \sigma_1(\sigma_1
W_2W_3W_4+W_1\sigma_2W_3W_4+W_1W_2\sigma_3W_4+W_1W_2W_3\sigma_4)\nonumber\\
&&\hspace{-8pt}=3\pi\sum_{\sigma_1,\sigma_2,\sigma_3,\sigma_4=\pm
1}\int_{\mathbb{T}^4} d k_1 d k_2 d k_3 d k_4(16
\omega_1\omega_2\omega_3\omega_4)^{-1}\delta
\big(\sum^4_{j=1}\sigma_j \omega_j\big)\delta
\big(\sum^4_{j=1}\sigma_j k_j\big)\nonumber\\
&&\hspace{14pt}\times W_1W_2W_3W_4 \big(\sum^4_{j=1}\sigma_j
W^{-1}_j\big)^2\,.
\end{eqnarray}
We conclude that the entropy production is non-negative and that
$dS(W)/dt=0$ if and only if $1/W$ is a positive collisional
invariant.

Vice versa, if $\psi$ is a collisional invariant and $\psi(k)\geq
0$, then setting $W(k)=1/\psi(k)$ one has
\begin{equation}\label{4.14}
\mathcal{C}(W)(k)=0\,.
\end{equation}
Thus stationary solutions are uniquely characterized by $1/W$ being
a positive collisional invariant. In brackets we remark that a
complete discussion would have to specify in which function space
one searches for collisional invariants.

Collisional invariants are studied in \cite{Sp4}, where the
following result is proved. $\omega$ is assumed to have no flat pieces
in the sense that the set $\{k \in \mathbb{T}^3\,|\,
\det \,\mathrm{Hess} \omega(k)  = 0\}$ has at most the dimension two.
Here Hess is the Hessian of $\omega$ as a $3\times3$ matrix.  Let $\psi:\mathbb{T}^3\to\mathbb{R}$ be
measurable, $\int_{\mathbb{T}^3} dk |\psi(k)| < \infty$, and a collisional invariant under pair collisions, in the
sense that (\ref{4.10}) holds Lebesgue almost surely. Then
$\psi$ is necessarily of the form
\begin{equation}\label{4.15}
\psi(k)=a+c\omega(k)
\end{equation}
with some constants $a,c\in\mathbb{R}$. By our previous discussion
this implies that the kinetic equation has a two-parameter family of
stationary solutions. If (\ref{4.15}) is inserted in the 3-phonons
merger (\ref{4.11}), then $a=0$ necessarily. To have thermal
equilibrium as only stationary solutions, there must be 3-phonons
mergers on a set of full dimension. A more physical mechanism would
be to include a cubic on-site potential, which appears naturally in
an expansion with respect to the anharmonicity, unless there are
special symmetries which would make this order to vanish. For the
cubic potential a collisional invariant must satisfy the 2-phonons
merger condition, which is (\ref{4.11}) upon dropping $k_3$. Again,
if there is a set of full dimension of 2-phonons mergers, then the
only stationary solutions are the ones corresponding to thermal
equilibrium.

Our discussion has interesting implications on the very widely
studied Fermi-Pasta-Ulam problem \cite{FPU}, see the special issue
\cite{C} at the occasion of its 50-th anniversary. Fermi, Pasta, and
Ulam considered chains of anharmonic oscillators and studied by a
numerical iteration scheme the relaxation to equipartition, i.e. to
thermal equilibrium. To their surprise, such an approach did not
take place, at least not on the time scale of the computation. One
explanation comes from the KAM theory which demonstrates that in
part the invariant tori of the linear dynamics persist under small
nonlinear perturbations. Thus a Lebesgue typical phase point may not
explore the full energy shell. A second explanation relies on the
fact that for a special choice of the nonlinearity the Hamiltonian
system remains integrable and admits propagating soliton solutions.
Also, for the FPU model there are special solutions, the breathers,
which may impede the relaxation to thermal equilibrium.

Kinetic theory operates in a different part of phase space. The
energy is proportional to the length of the chain and the
nonlinearity is weak. In this regime the phonon picture is precise.
Propagation is encoded by the dispersion relation $\omega$, while
the relaxation comes from the phonon collisions. For the FPU chain
one has $\omega(k)=\big(1-\cos(2\pi k)\big)^{1/2}$,
$k\in\mathbb{T}$. Thus, ignoring special $k$-values, phonons
propagate ballistically with non-zero speed. On the other hand, the
collision rule depends on the precise form of the nonlinearity. For
the FPU $\alpha$-chain the couplings are cubic as $(q_{j+1}-q_j)^3$.
In this case the collision term vanishes which signals poor
relaxation. For the FPU $\beta$-chain the couplings are quartic as
$(q_{j+1}-q_j)^4$. A detailed analysis shows that the energy current
correlation, as based on kinetic theory, has a slow decay as
$t^{-3/5}$ \cite{Per,LS}, a result with good numerical support \cite{LLP}. If instead one chooses as nonlinearity a
quartic on-site potential, as $q^4_j$, then the energy current
correlation decays exponentially \cite{ALS}. In this case kinetic
theory predicts a rapid convergence to equilibrium, as is well
confirmed by molecular dynamics simulations \cite{ALS,A}, also away
from the limiting kinetic regime.

\section{References}

\renewcommand{\refname}{}    
\vspace*{-26pt}              






\end{document}